\documentclass[a4paper,11pt]{article}

\usepackage{times}
\usepackage{mathptmx}
\usepackage{helvet}
\usepackage{courier}
\usepackage{amssymb}
\usepackage{pifont}
\usepackage{booktabs}
\usepackage{multirow}
\usepackage{eepic,epic}
\usepackage{epsfig}
\usepackage{graphicx}
\usepackage{graphics}
\usepackage{ifthen}
\usepackage{subfigure} 
\usepackage{paralist}
\usepackage[utf8]{inputenc}
\usepackage[T1]{fontenc}
\usepackage{array,xspace,tikz}
\usepackage{picture}
\usepackage{color}
\usepackage{mathrsfs}
\usepackage[left=4cm,right=4cm,%
]{geometry}

\newcommand{\game}{\ensuremath{\mathcal{G}}}
\newcommand{\gameNv}{\ensuremath{\game=(N,v)}}
\newcommand{\coal}{C\subseteq N}
\newcommand{\coalwoi}{\coal\smallsetminus\{i\}}

\newcommand{\bi}{\mathrm{Banzhaf}}
\newcommand{\shi}{\mathrm{ShapleyShubik}}

\newcommand{\bm}{\textsc{BeneficialMerge}}

\newcommand{\bs}{\textsc{BeneficialSplit}}

\newcommand{\sat}{\textsc{SAT}}
\newcommand{\threesat}{\textsc{3-SAT}}

\newcommand{\partition}{\textsc{Partition}}
\newcommand{\subsetsum}{\textsc{SubsetSum}}

\newcommand{\xthreeclong}{\rm \mbox{\sc{}ExactCoverBy3-Sets}}

\newcommand{\eq}{\enspace=\enspace}

\newcommand{\littlep}{\mathrm{p}}
\newcommand{\manyone}{\ensuremath{\mathrm{\,\leq_{\rm m}^{{\littlep}}\,}}}
\newcommand{\manyonetext}{\ensuremath{\mathrm{\leq_{\rm m}^{{\littlep}}}}}

\usepackage{amsmath,amsfonts,amssymb,nicefrac}

\usepackage{algorithm,algorithmic,multicol}

\usepackage{tabularx,multirow}

\usepackage{graphicx}

\usepackage{subfigure, color, epsfig}

\newcolumntype{L}{>{\raggedright\arraybackslash}X}
\newcolumntype{R}{>{\raggedleft\arraybackslash}X}
\newcolumntype{C}{>{\centering\arraybackslash}X}

\newcommand{\EP}[3]{
\smallskip
\begin{center}
{\small 
\begin{tabularx}{0.98\columnwidth}{ll}
\toprule
\multicolumn{2}{c}{\textsc{#1}} \\
\midrule
{\bf Given:}   & \parbox[t]{0.84\columnwidth}{#2\vspace*{1mm}}  \\
{\bf Question:}& \parbox[t]{0.84\columnwidth}{#3\vspace*{.5mm}} \\ 
\bottomrule
\end{tabularx}
}
\end{center}
\smallskip
}

\newcommand\qedblob{\mbox{\ding{113}}}

\def\literalqed{{\ \nolinebreak\hfill\mbox{\qedblob\quad}}}

\newenvironment{proofs}{\noindent{\sc
Proof.}}{\literalqed\bigskip}

\newcommand{\p}{\ensuremath{\mathrm{P}}}
\newcommand{\np}{\ensuremath{\mathrm{NP}}}

\newcommand{\pp}{\ensuremath{\mathrm{PP}}}
\newcommand{\ph}{\ensuremath{\mathrm{PH}}}

\newcommand{\sharpp}{\ensuremath{\mathrm{\#P}}}

\newcommand{\condition}{\,|\:}

\newcommand{\reals}{\mathbb{R}}

\newtheorem{theorem}{Theorem}[section]

\newtheorem{corollary}[theorem]{Corollary}

\newtheorem{lemma}[theorem]{Lemma}

\title{False-Name Manipulation in Weighted Voting Games is Hard
  for Probabilistic Polynomial Time
}

\author{Anja Rey and J\"{o}rg Rothe\\
Institut f\"{u}r Informatik\\
Heinrich-Heine-Universit\"{a}t D\"{u}sseldorf\\
40225 D\"{u}sseldorf, Germany
}
\date{\today}

\begin{document}

\maketitle

\begin{abstract}
  False-name manipulation refers to the question of whether a player
  in a weighted voting game can increase her power by splitting into
  several players and distributing her weight among these false
  identities.  Analogously to this splitting problem, the beneficial
  merging problem asks whether a coalition of players can increase
  their power in a weighted voting game by merging their weights.
  Aziz et al.~\cite{azi-bac-elk-pat:j:false-name-manipulations-wvg}
  analyze the problem of
  whether merging or
  splitting players in weighted voting games is beneficial in terms of
  the Shapley--Shubik and the normalized Banzhaf index,
  and so do
  Rey and Rothe~\cite{rey-rot:c:merging-splitting-banzhaf}
  for the probabilistic Banzhaf
  index.  All these results provide merely
  NP-hardness
  lower bounds for these
  problems, leaving the question about their exact complexity open.
  For the Shapley--Shubik and the probabilistic Banzhaf index, we
  raise these lower bounds to hardness for $\pp$, ``probabilistic
  polynomial time,'' and provide matching upper bounds for beneficial
  merging and, whenever the number of false identities is fixed, also
  for beneficial splitting, thus resolving
  previous conjectures in the affirmative.  It follows from our
  results that beneficial merging and splitting for these two power
  indices cannot be solved in $\np$, unless the polynomial hierarchy
  collapses, which is considered highly unlikely.
\end{abstract}

\section{Introduction}
\label{sec:introduction}

Weighted voting games are an important class of succinctly
representable, simple games.  They can be used to model cooperation
among players in scenarios where each player is assigned a weight, and
a coalition of players wins if and only if their joint weight meets or
exceeds a given quota.  Typical real-world applications of weighted
voting games include decision-making in legislative bodies (e.g.,
parliamentary voting) and shareholder voting (see the book by
Chalkiadakis et
al.~\cite{cha-elk-woo:b:computational-aspects-of-cooperative-game-theory}
for further concrete applications and literature pointers).  In
particular, the algorithmic and complexity-theoretic properties of
problems related to weighted voting have been studied in depth, see,
e.g., the work of Elkind et
al.~\cite{elk-cha-jen:c:coalition-structures-in-WVGs,elk-gol-gol-woo:j:complexity-of-weighted-threshold-games},
Bachrach et
al.~\cite{bac-elk-mei-pas-zuc-rot-ros:c:cost-of-stability}, Zuckerman
et al.~\cite{zuc-fal-bac-elk:c:manipulating-quote-in-wvg},
and~\cite{cha-elk-woo:b:computational-aspects-of-cooperative-game-theory}
for an overview.

Bachrach and Elkind~\cite{bac-elk:c:false-name-manipulation} were the
first to study false-name manipulation in weighted voting games: Is it
possible for a player to increase her power by splitting into several
players and distributing her weight among these false identities?
Relatedly, is it possible for two or more players to increase their
power in a weighted voting game by merging their weights?  The most
prominent measures of a player's power, or influence, in a weighted
voting game are the Shapley--Shubik and Banzhaf power indices.
Merging and extending the results
of~\cite{bac-elk:c:false-name-manipulation}
and~\cite{azi-pat:c:false-name-manipulation}, Aziz et
al.~\cite{azi-bac-elk-pat:j:false-name-manipulations-wvg} in
particular
study the problem of
whether merging or
splitting players in weighted voting games is beneficial in terms of
the Shapley--Shubik
index~\cite{sha:b:shapley-value,sha-shu:j:shapley-shubik-index} and
the normalized Banzhaf index~\cite{ban:j:weighted-voting-doesnt-work}
(see Section~\ref{sec:preliminaries} for formal definitions).  Rey and
Rothe~\cite{rey-rot:c:merging-splitting-banzhaf} extend this study
for the probabilistic Banzhaf
index proposed by Dubey and Shapley~\cite{dub-sha:j:banzhaf}.  All
these results, however, provide merely $\np$-hardness lower bounds.
Aziz et al.~\cite[Remark~13 on
p.~72]{azi-bac-elk-pat:j:false-name-manipulations-wvg} note that ``it
is quite possible that our problems are not in~$\np$'' (and thus are
not $\np$-complete).  Faliszewski and
Hemaspaandra~\cite{fal-hem:j:power-index-comparison} provide the best
known upper bound for the beneficial merging problem with respect to
the Shapley--Shubik index: It is contained in the class~$\pp$,
``probabilistic polynomial time,'' which is considered to be by far a
larger class than~$\np$, and they conjecture that this problem is
$\pp$-complete.  Rey and
Rothe~\cite{rey-rot:c:merging-splitting-banzhaf} observe that the same
arguments give a $\pp$ upper bound also for beneficial merging
in terms of the probabilistic Banzhaf index, and they
conjecture $\pp$-completeness
as
well.\footnote{They also note that the same arguments cannot be
  transferred immediately to the corresponding problem for the
  normalized Banzhaf index.}

We resolve these conjectures in the affirmative by
proving that beneficial merging and splitting 
(for any fixed number of
false identities)
are $\pp$-complete problems both
for the Shapley--Shubik and the probabilistic Banzhaf index.
Beneficial splitting in general (i.e., for an unbounded number of
false identities) belongs to $\np^{\pp}$
and is $\pp$-hard for the same two indices.  Thus,
none of these six problems can be in $\np$, unless the polynomial
hierarchy collapses to its first level, which is considered highly
unlikely.

\section{Preliminaries}
\label{sec:preliminaries}

We will need the following concepts from cooperative game theory (see,
e.g., the textbook by Chalkiadakis et
al.~\cite{cha-elk-woo:b:computational-aspects-of-cooperative-game-theory}).
A \emph{coalitional game with transferable utilities}, $\gameNv$,
consists of a set $N=\{1,\dots, n\}$ of \emph{players} (or,
synonymously, \emph{agents}) and a \emph{coalitional function} $v:
\mathfrak{P}(N)\rightarrow \reals$ with $v(\emptyset) = 0$, where
$\mathfrak{P}(N)$ denotes the power set of~$N$.  $\game$ is
\emph{monotonic} if $v(B) \leq v(C)$ whenever $B \subseteq C$ for
coalitions $B,C \subseteq N$, and it is \emph{simple} if it is
monotonic and $v: \mathfrak{P}(N)\rightarrow \{0,1\}$, that is, $v$
maps each coalition $C\subseteq N$ to a value that indicates whether
$C$ \emph{wins} (i.e., $v(C) = 1$) or \emph{loses} (i.e., $v(C) = 0$),
where we require that the grand coalition $N$ is always winning.  The
\emph{probabilistic Banzhaf power index of a player $i\in N$ in a
  simple game~$\game$} (see~\cite{dub-sha:j:banzhaf}) is defined by
\begin{equation}
\label{eq:banzhaf}
  \bi(\game,i) \eq \frac{1}{2^{n-1}}
  \sum_{C\subseteq N \smallsetminus \{i\}}
  \left(v(C\cup \{i\}) - v(C)\right).
\end{equation}
Intuitively, this index measures the power of player $i$ in terms of
the probability such that $i$ turns a losing coalition $C\subseteq N
\smallsetminus \{i\}$ into a winning coalition by joining it, and
therefore is \emph{pivotal} for the success of~$C$.  (For comparison,
the \emph{normalized Banzhaf index of $i$ in~$\game$}
defined by Banzhaf~\cite{ban:j:weighted-voting-doesnt-work},
who rediscovered a notion originally introduced by
Penrose~\cite{pen:j:banzhaf-index},
is obtained by
dividing the \emph{raw Banzhaf index of $i$ in~$\game$}, which is the
term $\sum_{C\subseteq N \smallsetminus \{i\}} \left(v(C\cup \{i\}) -
  v(C)\right)$ in~(\ref{eq:banzhaf}), not by $2^{n-1}$, but by the sum
of the raw Banzhaf indices of all players in~$\game$; see
\cite{dub-sha:j:banzhaf,fel-mac:j:voting-power-measurement-misreinvention,rey-rot:c:merging-splitting-banzhaf} for a
discussion of the differences between these two power indices.)

Unlike the Banzhaf indices, the \emph{Shapley--Shubik index of $i$
  in~$\game$} takes the order into account in which players enter
coalitions and is defined by
\[
\shi(\game,i) \eq \frac{1}{n!}\sum_{\coalwoi}
\|C\|!\cdot(n-1-\|C\|)!\cdot \left(v(C\cup \{i\}) - v(C)\right).
\]
Since the number of coalitions is exponential in the number of
players, specifying coalitional games by listing all values of their
coalitional function would require exponential space.  For algorithmic
purposes, however, it is important that these games can be represented
succinctly.  Simple games can be compactly represented by weighted
voting games.  A \emph{weighted voting game} (\emph{WVG})
$\game=(w_1,\dots,w_n;\,q)$ consists of nonnegative integer
weights~$w_i$, $1\leq i\leq n$, and a quota~$q$, where $w_i$ is the
$i$th player's weight.  For each coalition $C\subseteq N$, letting
$w(C)$ denote $\sum_{i\in C} w_i$, $C$ wins
if $w(C) \geq q$, and it loses
otherwise.  Requiring the quota to satisfy $0 < q \leq w(N)$ ensures
that the empty coalition loses and the grand coalition wins.  Weighted
voting games have been intensely studied from a computational
complexity point of view (see, e.g.,
\cite{elk-cha-jen:c:coalition-structures-in-WVGs,elk-gol-gol-woo:j:complexity-of-weighted-threshold-games,bac-elk-mei-pas-zuc-rot-ros:c:cost-of-stability,zuc-fal-bac-elk:c:manipulating-quote-in-wvg}
and
\cite[Chapter~4]{cha-elk-woo:b:computational-aspects-of-cooperative-game-theory}
for an overview).

Aziz et al.~\cite{azi-bac-elk-pat:j:false-name-manipulations-wvg}
introduce the merging and splitting operations for WVGs. %
We use the following notation.
Given a WVG $\game=(w_1,\dots,w_n;\,q)$ and a nonempty\footnote{We omit the
empty coalition, since this would slightly change the idea of the problem.}
coalition
$S\subseteq\{1,\dots,n\}$, let $\game_{\&S}=(w(S),w_{j_1},
\dots,w_{j_{n-\|S\|}};\,q)$ with $\{j_1, \dots, j_{n-\|S\|}\} =
N\smallsetminus S$ denote the new WVG in which the players in $S$ have
been merged into one new player of weight~$w(S)$.
Similarly, given a WVG $\game= (w_1,\dots,w_n;\,q)$, a player~$i$, and
an integer $m \geq 2$, define the set of WVGs
\[
\game_{i\div m} =
(w_1,\dots,w_{i-1},w_{i_1},\dots,w_{i_m},w_{i+1},\dots,w_n;\,q)
\]
in which $i$ with weight~$w_i$ is split into $m$ new players
$i_1,\dots,i_m$ with weights $w_{i_1},\dots,w_{i_m}$ such that
$\sum_{j=1}^m w_{i_j} = w_i$.  (Note that there is a \emph{set} of
such WVGs $\game_{i\div m}$, since there might be several possibilities of
distributing $i$'s weight $w_i$ to the new players $i_1,\dots,i_m$
satisfying $\sum_{j=1}^m w_{i_j} = w_i$.)
For a power index \textsc{PI}, the beneficial merging and splitting
problems are defined as follows.

\EP{\textsc{PI}-$\bm$}
{A WVG $\game=(w_1,\dots,w_n;\,q)$ and a nonempty
 coalition $S\subseteq\{1,\dots,n\}$.}
{Is it true that
 $\textsc{PI}(\game_{\&S},1) > \sum_{i\in S} \textsc{PI}(\game,i)$?}

We distinguish between two splitting problems: In the first problem,
the number $m$ of false identities some player splits into is not
part of the given problem instance (rather, the problem itself is
parameterized by~$m$), whereas $m$ is given in the instance for the
second problem.  (This distinction wouldn't make sense for
beneficial merging.)

\EP{\textsc{PI}-$m$-$\bs$}
{A WVG $\game=(w_1,\dots,w_n;\,q)$ and a player~$i$.}
{Is it possible to split $i$ into $m$ new players
 $i_1,\dots,i_m$ with weights $w_{i_1},\dots,w_{i_m}$ satisfying
 $\sum_{j=1}^m w_{i_j} = w_i$ such that in this new WVG $\game_{i\div m}$,
 it holds that $\sum_{j=1}^m \textsc{PI}(\game_{i\div m},i_j) >
 \textsc{PI}(\game,i)$?}

\EP{\textsc{PI}-$\bs$}
{A WVG $\game=(w_1,\dots,w_n;\,q)$, a player~$i$, and
 an integer $m \geq 2$.}
{Is it possible to split $i$ into $m$ new players
 $i_1,\dots,i_m$ with weights $w_{i_1},\dots,w_{i_m}$ satisfying
 $\sum_{j=1}^m w_{i_j} = w_i$ such that in this new WVG $\game_{i\div m}$,
 it holds that $\sum_{j=1}^m \textsc{PI}(\game_{i\div m},i_j) >
 \textsc{PI}(\game,i)$?}

The goal of this paper is to classify these problems in terms of
their complexity for both the Shapley--Shubik and the probabilistic
Banzhaf index.  We assume that the reader is familiar with the basic
complexity-theoretic concepts such as the complexity classes $\p$ and
$\np$, the polynomial-time many-one reducibility, denoted
by~$\manyonetext$, and the notions of hardness and completeness with
respect to~$\manyonetext$ (see, e.g., the textbook by
Papadimitriou~\cite{pap:b:complexity}).
Valiant~\cite{val:j:permanent} introduced $\sharpp$ as the class of
functions that give the number of solutions of the instances of $\np$
problems.  For a decision problem $A \in \np$, we denote this function
by~$\#A$.  For example, if $\sat$ is the satisfiability problem from
propositional logic, then $\#\sat$ denotes the function mapping any
boolean formula $\varphi$ to the number of truth assignments
satisfying~$\varphi$.  There are various notions of reducibility
between functional problems in $\sharpp$
(see~\cite{fal-hem:j:power-index-comparison} for an overview,
literature pointers, and discussion).  Here, we need only the most
restrictive one: We say \emph{a function $f$ parsimoniously reduces to
  a function $g$} if there exists a polynomial-time computable
function $h$ such that for each input~$x$, $f(x)=g(h(x))$.  That is,
for functional problems $f,g \in \sharpp$, a parsimonious reduction
$h$ from $f$ to $g$ transfers each instance $x$ of $f$ into an
instance $h(x)$ of $g$ such that $f(x)$ and $g(h(x))$ have the same
number of solutions.  We say that $g$ is $\sharpp$-parsimonious-hard
if every $f \in \sharpp$ parsimoniously reduces to~$g$.  We say that
$g$ is $\sharpp$-parsimonious-complete if $g$ is in $\sharpp$ and
$\sharpp$-parsimonious-hard.  It is known that, given a WVG $\game$
and a player~$i$, computing the raw Banzhaf index is
$\sharpp$-parsimonious-complete~\cite{pra-kel:j:voting}, whereas
computing the raw Shapley--Shubik index is
not~\cite{fal-hem:j:power-index-comparison},
although it, of course, is in $\sharpp$ as well.

Gill~\cite{gil:j:probabilistic-tms} introduced the class $\pp$
(``probabilistic polynomial time'') that contains all decision
problems $X$ for which there exist a function $f \in \sharpp$ and a
polynomial $p$ such that for all instances~$x$, $x \in X$ if and only
if $f(x) \geq 2^{p(|x|)-1}$.  It is easy to see that $\np \subseteq
\pp$; in fact, $\pp$ is considered to be by far a larger class
than~$\np$, due to Toda's theorem~\cite{tod:j:pp-ph}: $\pp$ is at
least as hard (in terms of polynomial-time Turing reductions) as any
problem in the polynomial hierarchy (i.e., $\ph \subseteq \p^{\pp}$).
$\np^{\pp}$, the second level of Wagner's counting
hierarchy~\cite{wag:j:succinct}, is the class of problems solvable by
an $\np$ machine with access to a $\pp$ oracle; Mundhenk et
al.~\cite{mun-gol-lus-all:j:finite-horizon-markov-decision-process-complexity}
identified $\np^{\pp}$-complete problems related to finite-horizon
Markov decision processes.

\section{Beneficial Merging and Splitting is PP-Hard}
\label{sec:pp-completeness}

In this section we prove that beneficial merging and splitting is
$\pp$-hard, and we provide matching upper bounds for beneficial
merging and splitting (for any fixed number of false identities)
both for the Shapley--Shubik and the probabilistic
Banzhaf index.  We start with the latter.

\subsection{The Probabilistic Banzhaf Power Index}
\label{sec:pp-completeness-banzhaf}

We will use the following result due to Faliszewski and
Hemaspaandra~\cite[Lemma~2.3]{fal-hem:j:power-index-comparison}.

\begin{lemma}[Faliszewski and
  Hemaspaandra~\cite{fal-hem:j:power-index-comparison}]
\label{lem:compare-pp-complete}
\quad
  Let $F$ be a $\sharpp$-parsimonious-com\-plete function.  The problem
  $\textsc{Compare}\hbox{-}F = \{(x,y) \condition F(x) > F(y)\}$ is
  $\pp$-complete.
\end{lemma}

The well-known $\np$-complete problem \textsc{SubsetSum} (which is a
special variant of the \textsc{Knapsack} problem) asks, given a sequence
$(a_1,\dots,a_n)$ of positive integers and a positive integer~$q$, do
there exist $x_1,\dots,x_n \in \{0,1\}$ such that $\sum_{i=1}^{n} x_i
a_i = q$?  It is known that $\#\textsc{SubsetSum}$ is
$\sharpp$-parsimonious-complete (see, e.g., the textbook by
\cite{pap:b:complexity}
for parsimonious reductions from $\#\threesat$ via
$\#\textsc{ExactCoverBy3-Sets}$ to
$\#\subsetsum$).  Hence, by Lemma~\ref{lem:compare-pp-complete}, we
have the following.

\begin{corollary}
\label{cor:compare-subsetsum-pp-complete}
 \textsc{Compare}-\textsc{\#SubsetSum} is $\pp$-complete.
\end{corollary}

Our goal is to
$\manyonetext$-reduce
\textsc{Compare}-\textsc{\#SubsetSum} to
$\bi$-\textsc{Bene}-\textsc{ficialMerge}.  However, to make this reduction work,
it will be useful to consider two restricted variants of
\textsc{Compare}-\textsc{\#SubsetSum}, which we denote by
\textsc{Compare}-\textsc{\#Subset}-\textsc{Sum}-\textsc{R} and
\textsc{Compare}-\textsc{\#SubsetSum}-\textsc{RR}, show their
$\pp$-hardness, and then reduce
\textsc{Compare}-\textsc{\#SubsetSum}-\textsc{RR} to
the problem $\bi$-\textsc{BeneficialMerge}.
This will be done in
Lemmas~\ref{lem:restr-compare-subsetsum-pp-complete}
and~\ref{lem:even-more-restr-compare-subsetsum-pp-complete} and in
Theorem~\ref{thm:bmb-pp-complete}.  In all restricted variants of
\textsc{Compare}-\textsc{\#SubsetSum} we may assume, without loss of
generality, that the target value $q$ in the related 
\textsc{\#SubsetSum} instances
$((a_1,\dots,a_n),q)$ satisfies $1\leq q \leq \alpha -1$, where
$\alpha =\sum_{i=1}^n a_i$, such that $\#\subsetsum$ remains
$\sharpp$-parsimonious-complete.

\EP{\textsc{Compare}-\textsc{\#SubsetSum}-\textsc{R}}
{A sequence $A=(a_1,\dots,a_n)$ of positive integers
 and two positive integers $q_1$ and $q_2$ with
 $1\leq q_1,q_2\leq \alpha -1$, where $\alpha =\sum_{i=1}^na_i$.}
{Is the number of 
  subsequences of $A$ summing up to $q_1$ greater than the number of
  subsequences of $A$ summing up to~$q_2$, that is, does it hold that
$\#\subsetsum((a_1,\dots,a_n),q_1) > \#\subsetsum((a_1,\dots,a_n),q_2)$?
}

\begin{lemma}
\label{lem:restr-compare-subsetsum-pp-complete}
$\textsc{Compare}\hbox{-}\textsc{\#SubsetSum} \manyone
\textsc{Compare}\hbox{-}\textsc{\#SubsetSum}\hbox{-}\textsc{R}$.
\end{lemma}

\begin{proofs}
 Given an instance $(X,Y)$ of
 \textsc{Compare}\hbox{-}\textsc{\#SubsetSum},
 $X=((x_1,\dots,x_{m}),q_x)$ and $Y=((y_1,\dots,y_{n}),q_y)$,
 construct a
 \textsc{Compare}\hbox{-}\textsc{\#SubsetSum}\hbox{-}\textsc{R}
 instance $(A,q_1,q_2)$ as follows.
 Let $\alpha =\sum_{i=1}^{m}x_i$ and define
 $A=(x_1,\dots,x_{m},2\alpha y_1,\dots,2\alpha y_{n})$,
 $q_1=q_x$, and $q_2=2\alpha q_y$.
 This construction can obviously be achieved in polynomial time.
 
 It holds that integers from $A$ can only sum up to $q_x<\alpha -1$
 if they do not contain multiples of $2\alpha $, thus
 $\#\subsetsum(A,q_1)
 =\#\subsetsum((x_1,\dots,x_{m}),q_x)$.
 On the other hand, $q_2$ can only be obtained by multiples of $2\alpha $,
 since $\sum_{i=1}^{m}x_i=\alpha $ is too small.
 Thus, it holds that $\#\subsetsum(A,q_2)
 =\#\subsetsum((y_1,\dots,y_{n}),q_y)$.
 It follows that $(X,Y)$ is in
 \textsc{Compare}\hbox{-}\textsc{\#SubsetSum} if and only if
 $(A,q_1,q_2)$ is in
\textsc{Compare}\hbox{-}\textsc{\#SubsetSum}\hbox{-}\textsc{R}.~\end{proofs}

In order to perform the next step, we need to ensure that
all integers in a
\textsc{Compare}\hbox{-}\textsc{\#Sub}-\textsc{setSum}\hbox{-}\textsc{R}
instance are divisible by~$8$.
This can easily be achieved, since for a given instance
$((a_1,\dots,a_n),q_1,q_2)$, we can multiply each integer
by $8$, obtaining $((8a_1,\dots,8a_n),8q_1,8q_2)$
without changing the number of
solutions for both related
\textsc{SubsetSum} instances.
Thus, from now on, without loss of generality,
we assume that for a given
\textsc{Compare}\hbox{-}\textsc{\#Subset}-\textsc{Sum}\hbox{-}\textsc{R}
instance $((a_1,\dots,a_n),q_1,q_2)$, it holds that
$a_i,q_j\equiv 0\bmod 8$ for $1\leq i\leq n$ and $j\in\{1,2\}$.

Now, we consider our even more restricted variant of this problem.

\EP{\textsc{Compare}\hbox{-}\textsc{\#SubsetSum}\hbox{-}\textsc{RR}}
{A sequence $A=(a_1,\dots,a_n)$ of positive integers.}
{Is the number of
 subsequences of $A$ summing up to $\left(\nicefrac{\alpha}{2}\right)-2$
 greater than the number of
 subsequences of $A$ summing up to
 $\left(\nicefrac{\alpha }{2}\right)-1$, i.e.,
 $\#\subsetsum((a_1,\dots,a_n),(\nicefrac{\alpha }{2})-2)
 >\#\subsetsum((a_1,\dots,a_n),(\nicefrac{\alpha }{2})-1)$, where
 $\alpha =\sum_{i=1}^na_i$?}

\begin{lemma}
\label{lem:even-more-restr-compare-subsetsum-pp-complete}
$\textsc{Compare}\hbox{-}\textsc{\#SubsetSum}\hbox{-}\textsc{R} \manyone
\textsc{Compare}\hbox{-}\textsc{\#SubsetSum}\hbox{-}\textsc{RR}$.
\end{lemma}

\begin{proofs}
  Given an instance $(A,q_1,q_2)$ of
  \textsc{Compare}\hbox{-}\textsc{\#SubsetSum}\hbox{-}\textsc{R},
  where we assume that $A=(a_1,\dots,a_n)$, $q_1$, and $q_2$ satisfy
  $a_i,q_j\equiv 0\bmod 8$ for $1\leq i\leq n$ and $j\in\{1,2\}$,
  we construct 
  an instance $B$ of
  \textsc{Compare}\hbox{-}\textsc{\#SubsetSum}\hbox{-}\textsc{RR} as
  follows.  (This reduction is inspired by the standard reduction
  from $\subsetsum$ to $\partition$ due to Karp~\cite{kar:b:reducibilities}.)

  Letting $\alpha =\sum_{i=1}^n a_i$, define
 \[
  B=(a_1,\dots,a_n, 2\alpha -q_1, 2\alpha +1-q_2, 2\alpha +3+q_1+q_2, 3\alpha ).
 \]
 This instance can obviously be constructed in polynomial time.
 Observe that
  \[
  T=\left(\sum_{i=1}^n a_i\right)+(2\alpha -q_1)+(2\alpha
   +1-q_2) +(2\alpha +3+q_1+q_2)+3\alpha =10\alpha +4,
  \]
  and therefore, $\left(\nicefrac{T}{2}\right)-2=5\alpha $ and
  $\left(\nicefrac{T}{2}\right)-1=5\alpha +1$.
 
 We show that $(A,q_1,q_2)$
 is in \textsc{Compare}\hbox{-}\textsc{\#SubsetSum}\hbox{-}\textsc{R}
 if and only if the constructed instance $B$
 is in \textsc{Compare}\hbox{-}\textsc{\#SubsetSum}\hbox{-}\textsc{RR}.
 
 First, we examine which subsequences of $B$ sum up to $5\alpha$.
 Consider the following cases: If
 $3\alpha $ is added, $2\alpha +3+q_1+q_2$ cannot be added, as it
 would be too large.  Also, $2\alpha +1-q_2$ cannot be added, leading
 to an odd sum.  So, $2\alpha -q_1$ has to be added, as the remaining
 $\alpha $ are too small.  Since $3\alpha +2\alpha -q_1=5\alpha -q_1$,
 $5\alpha $ can be achieved by adding some $a_i$'s 
 if and only if there
 exists a subset $A'\subseteq\{1,\dots,n\}$ such that $\sum_{i\in
   A'}a_i=q_1$ (i.e., $A'$ is a
 solution
 of the \textsc{SubsetSum} instance $(A,q_1)$).  If $3\alpha$ is not
 added, but $2\alpha +3+q_1+q_2$, an even number can only be achieved
 by adding $2\alpha +1-q_1$ such that $\alpha -4-q_1$ remain. So,
 $2\alpha -q_1$ is too large, while no subsequence
 of $A$ sums up to
 $\alpha -4-q_1$, because of the assumption of divisibility by $8$.
 If neither $3\alpha $ nor $2\alpha +3+q_1+q_2$ are added, the
 remaining $5\alpha +1-q_1-q_2$ are too small.  Thus, the only
 possibility to obtain $5\alpha$
 is to find a subsequence of $A$ adding up to~$q_1$.  Thus,
 $\#\subsetsum(A,q_1)=\#\subsetsum(B,5\alpha )$.
 
 Second, for similar reasons, a sum of $5\alpha +1$ can only be achieved by
 adding $3\alpha +(2\alpha +1-q_2)$ and a term $\sum_{i\in A'}a_i$, where
 $A'$ is a subset of $\{1,\dots,n\}$ such that $\sum_{i\in
   A'}a_i=q_2$.  Hence,
 $\#\subsetsum(A,q_2)=\#\subsetsum(B,5\alpha +1)$.
 
 Thus, the relation
 $\#\subsetsum(A,q_1)>\#\subsetsum(A,q_2)$
 holds if and only if
 $\#\textsc{Sub}$-$textsc{setsum}(B,5\alpha )>\#\subsetsum(B,5\alpha +1)$,
 which completes the proof.~\end{proofs}

We now are ready to prove the main theorem of this section.

\begin{theorem}
\label{thm:bmb-pp-complete}
$\bi$-\textsc{BeneficialMerge} is $\pp$-complete, even if only three
players of equal weight merge.
\end{theorem}

\begin{proofs}
 Membership of $\bi$-\textsc{BeneficialMerge} in $\pp$ has already
 been observed in~\cite[Theorem~3]{rey-rot:c:merging-splitting-banzhaf}.
 It follows from the fact that the raw Banzhaf index is in~$\sharpp$ and
 that $\sharpp$ is closed under addition
 and multiplication by two,\footnote{Again, note
 that this idea cannot be transferred straightforwardly to the normalized
 Banzhaf index, since in different games the indices have possibly different
 denominators, not only different by a factor of some power of two,
 as is the case for the probabilistic Banzhaf index.}
 and, furthermore, since comparing the values of two $\sharpp$
 functions on two (possibly different) inputs 
 reduces to a $\pp$-complete problem.  This technique (which
 was proposed by Faliszewski and
 Hemaspaandra~\cite{fal-hem:j:power-index-comparison}
 and applies their Lemma 2.10)
 works, since $\pp$ is closed under $\manyone$-reducibility.

  We show $\pp$-hardness of $\bi$-\textsc{BeneficialMerge} by means of
  a $\manyonetext$-reduction from
  \textsc{Compare}\hbox{-}\textsc{\#SubsetSum}\hbox{-}\textsc{RR},
  which is $\pp$-hard by
  Corollary~\ref{cor:compare-subsetsum-pp-complete} via
  Lemmas~\ref{lem:restr-compare-subsetsum-pp-complete}
  and~\ref{lem:even-more-restr-compare-subsetsum-pp-complete}. Our
  construction %
  is inspired by
  the $\np$-hardness results by Aziz
  et al.~\cite{azi-pat:c:false-name-manipulation} and Rey and
  Rothe~\cite{rey-rot:c:merging-splitting-banzhaf}. 

  Given an instance $A=(a_1,\dots,a_n)$ of
  \textsc{Compare}\hbox{-}\textsc{\#SubsetSum}\hbox{-}\textsc{RR},
  construct the following instance for $\bi$-\textsc{BeneficialMerge}.
  Let $\alpha =\sum_{i=1}^na_i$.  Define the WVG
\[
\game=(2a_1,\dots, 2a_n,1,1,1,1;\,\alpha ),
\]
and let the merging coalition be $S = \{n+2,n+3,n+4\}$.
  Letting $N = \{1,\dots,n\}$, it holds that
\begin{eqnarray*}
\bi(\game,n+2)
 & = & 
\frac{1}{2^{n+3}}
    \left\|\left\{C\subseteq\{1,\dots,n+1,n+3,n+4\} ~\middle|~
\sum_{i\in C}w_i=\alpha -1\right\}\right\|
\end{eqnarray*}
\vspace*{-6mm}
\begin{eqnarray}
   & \hspace*{-6mm} = & 
\label{eq:first}
\frac{1}{2^{n+3}}\left(
    \left\|\left\{A'\subseteq N  ~\middle|~ 
\sum_{i\in A'}2a_i=\alpha -1\right\}\right\|
 + 3\cdot\left\|\left\{A'\subseteq N ~\middle|~
 1+\sum_{i\in A'}2a_i=\alpha -1\right\}\right\| \right.
\\ & \hspace*{-6mm}   & 
\label{eq:second}
 \left.  
  + 3\cdot\left\|\left\{A'\subseteq N ~\middle|~
   2+\sum_{i\in A'}2a_i=\alpha -1\right\}\right\|
  +\left\|\left\{A'\subseteq N ~\middle|~ 3+\sum_{i\in A'}2a_i
 =\alpha -1\right\}\right\|
    \right)
\\ & \hspace*{-6mm} = & 
\frac{1}{2^{n+3}}\left(
    3\cdot\left\|\left\{A'\subseteq N ~\middle|~
 \sum_{i\in A'}2a_i=\alpha -2\right\}\right\|
  +\left\|\left\{A'\subseteq N ~\middle|~
 \sum_{i\in A'}2a_i=\alpha -4\right\}\right\|
    \right),
\nonumber
\end{eqnarray}
 since the $2a_i$'s can only add up to an even number.
 The first of the four sets in~(\ref{eq:first}) and~(\ref{eq:second})
 refers to those coalitions that do not contain any of the players $n+1$,
 $n+3$, and $n+4$; the second, third, and fourth set
 in~(\ref{eq:first}) and~(\ref{eq:second})
 refers to those coalitions containing either one, two, or three of them,
 respectively.
 Since they all have the same weight, players $n+3$ and $n+4$ have the same
 probabilistic Banzhaf index as player $n+2$.
 
 Furthermore, the new game after merging is
 $\game_{\&\{n+2,n+3,n+4\}}=(3,2a_1,\dots 2a_n,1;\,\alpha)$ and,
 similarly to above, the Banzhaf index of the first player is
 calculated as follows:
 \begin{eqnarray*}
  \lefteqn{\bi\left(\game_{\&\{n+2,n+3,n+4\}},1\right)}
\\ & = & \frac{1}{2^{n+1}}
    \left\|\left\{C\subseteq\{2,\dots,n+2\}~\middle|~ \sum_{i\in C}w_i=\alpha-1\right\}\right\|
\\ & = & \frac{1}{2^{n+1}}\left(
    \left\|\left\{A'\subseteq N ~\middle|~ \sum_{i\in A'}2a_i\in\{\alpha-3,\alpha-2,\alpha-1\}\right\}\right\|
    \right.
\\ &   &
\left.
    +\left\|\left\{A'\subseteq N ~\middle|~ 1+\sum_{i\in A'}2a_i\in\{\alpha-3,\alpha-2,\alpha-1\}\right\}\right\|
    \right)
\\ & = & \frac{1}{2^{n+1}}\left(
    2\cdot\left\|\left\{A'\subseteq N ~\middle|~ \sum_{i\in A'}2a_i=\alpha-2\right\}\right\|
    +\left\|\left\{A'\subseteq N ~\middle|~ \sum_{i\in A'}2a_i=\alpha-4\right\}\right\|
    \right).    
 \end{eqnarray*}
 Altogether, it holds that
 \begin{eqnarray*}
 \lefteqn{\bi\left(\game_{\&\{n+2,n+3,n+4\}},1\right)\quad
 -\sum_{i\in\{n+2,n+3,n+4\}}\bi(\game,i)}
\\ & = & \frac{1}{2^{n+1}}\left(
    2\cdot\left\|\left\{A'\subseteq N ~\middle|~ \sum_{i\in A'}2a_i=\alpha-2\right\}\right\|
    +\left\|\left\{A'\subseteq N ~\middle|~ \sum_{i\in A'}2a_i=\alpha-4\right\}\right\|
    \right)
\\ &   &
    -\frac{3}{2^{n+3}}\left(
    3\cdot\left\|\left\{A'\subseteq N ~\middle|~ \sum_{i\in A'}2a_i=\alpha-2\right\}\right\|
    +\left\|\left\{A'\subseteq N ~\middle|~ \sum_{i\in A'}2a_i=\alpha-4\right\}\right\|
    \right)
\\ & = & \left(\frac{1}{2^{n+1}}\cdot2-\frac{3}{2^{n+3}}\cdot 3\right)
    \left\|\left\{A'\subseteq N ~\middle|~ \sum_{i\in A'}2a_i=\alpha-2\right\}\right\|
\\ &   &
    +\left(\frac{1}{2^{n+1}}-\frac{3}{2^{n+3}}\right)
    \left\|\left\{A'\subseteq N ~\middle|~ \sum_{i\in A'}2a_i=\alpha-4\right\}\right\|
\\ & = & -\frac{1}{2^{n+3}}\cdot
    \left\|\left\{A'\subseteq N ~\middle|~ \sum_{i\in A'}a_i=\frac{\alpha}{2}-1\right\}\right\|
    +\frac{1}{2^{n+3}}\cdot
    \left\|\left\{A'\subseteq N ~\middle|~ \sum_{i\in A'}a_i=\frac{\alpha}{2}-2\right\}\right\|,
 \end{eqnarray*}
 which is greater than zero if and only if
 $\|\{A'\subseteq N \condition \sum_{i\in A'}a_i=\left(\nicefrac{\alpha}{2}\right)-2\}\|$
 is greater than
 $\|\{A'\subseteq N \condition \sum_{i\in A'}a_i=\left(\nicefrac{\alpha}{2}\right)-1\}\|$,
 which in turn is the case if and only if the original instance $A$ is in
\textsc{Compare}\hbox{-}\textsc{\#SubsetSum}\hbox{-}\textsc{RR}.~\end{proofs}

It is known (see~\cite{rey-rot:c:merging-splitting-banzhaf}) that both
the beneficial merging problem for a coalition $S$ of size $2$ and the
beneficial splitting problem for $m=2$ false identities can trivially
be decided in polynomial time for the probabilistic Banzhaf index,
since the sum of power (in terms of this index) of two players is
always equal to the power of the player that is obtained by merging
these two players.  Analogously to the proof of
Theorem~\ref{thm:bmb-pp-complete}, it can be shown that the beneficial
splitting problem for a fixed number of at least three false
identities is $\pp$-complete.

Since we allow players with zero weight, we need another
simple fact for the analysis of the beneficial splitting problem to be
used in the proofs of Theorems~\ref{thm:mbsb-pp-complete}
and~\ref{thm:mbssh-pp-complete}.

\begin{lemma}
\label{lem:bi-0-player}
For both the probabilistic Banzhaf index and the Shapley--Shubik
index, given a weight\-ed voting game, adding a player with weight zero
does not change the original players' power indices, and the new
player's power index is zero.
\end{lemma}

The proof of Lemma~\ref{lem:bi-0-player} is straightforward and
therefore omitted.  We are now ready to prove
Theorem~\ref{thm:mbsb-pp-complete}, which states that
$\bi$-$m$-\textsc{BeneficialSplit} is $\pp$-complete for each $m\geq3$.

\begin{theorem}
  \label{thm:mbsb-pp-complete}
  $\bi$-$m$-\textsc{BeneficialSplit} is $\pp$-complete for each
  $m\geq3$.
\end{theorem}

\begin{proofs}
As already mentioned in the proof of
Theorem~\ref{thm:bmb-pp-complete}, comparing the values of two
$\sharpp$ functions on two (possibly different) inputs reduces to a
$\pp$-complete problem and thus is is in~$\pp$.  In particular, this
is true for the problem of comparing (sums of) 
probabilistic Banzhaf indices in possibly different WVGs, such as
testing whether
the sum of the new players' raw Banzhaf indices is greater than
$2^{m-1}$ times the raw Banzhaf index of the original player $i$
(which is equivalent to
``$\sum_{j=1}^m \bi(\game_{i\div m},i_j) > \bi(\game,i)$''
from the definition of $\bi$-$m$-\textsc{Bene}-\textsc{ficialSplit}),
where $i$ is split into $m$ new players $i_1,\dots,i_m$.

The main difference between the beneficial merging and splitting
problems is that before comparing the two $\sharpp$ functions
associated with beneficial splitting, one has to choose a right way of
distributing $i$'s weight among the $m$ false identities of~$i$.
Since $m$ is fixed, there are only polynomially many (specifically,
some number in $\mathscr{O}(w_i^m)$) ways of doing so, i.e., of
finding nonnegative integers $w_{i_1},\dots,w_{i_m}$ satisfying
$\sum_{j=1}^m w_{i_j} = w_i$.  Thus, this comparison can be done
in~$\pp$ for each such weight distribution.  As $\pp$ is closed under
union, $\bi$-$m$-\textsc{BeneficialSplit} is in~$\pp$.
 
In order to show $\pp$-hardness for
$\bi$-$3$-\textsc{BeneficialSplit}, we use the same techniques as in
Theorem~\ref{thm:bmb-pp-complete}, appropriately modified.

First, we slightly change the definition of
\textsc{Compare}\hbox{-}\textsc{\#SubsetSum}\hbox{-}\textsc{RR} by
switching $(\nicefrac{\alpha}{2})-2$ and $(\nicefrac{\alpha}{2})-1$.
The problem (call it 
\textsc{Compare}\hbox{-}\textsc{\#SubsetSum}\hbox{-}\textsc{\reflectbox{RR}})
of whether the number of subsequences of a given sequence
$A$ of positive integers summing up to $(\nicefrac{\alpha}{2})-1$ is
greater than the number of subsequences of $A$ summing up to
$(\nicefrac{\alpha}{2})-2$, is $\pp$-hard by the same proof as in
Lemma~\ref{lem:even-more-restr-compare-subsetsum-pp-complete} with the
roles of $q_1$ and $q_2$ exchanged.

Now, we reduce this problem to $\bi$-$3$-\textsc{BeneficialSplit} by
constructing the following instance of the beneficial splitting problem
from an instance
$A=(a_1,\dots,a_n)$ of the problem
\textsc{Compare}\hbox{-}\textsc{\#SubsetSum}\hbox{-}\textsc{\reflectbox{RR}}.
Let $\game=(2a_1,\dots,2a_n,1,3;\,\alpha)$,
where $\alpha=\sum_{j=1}^na_j$, and let $i=n+2$ be the player to be
split.  $\game$ is (apart from the order of players) equivalent to the
game obtained by merging in the proof of
Theorem~\ref{thm:bmb-pp-complete}.  Thus, letting
$N=\{1,\dots,n\}$, $\bi(\game,n+2)$ equals
 \[
 \frac{1}{2^{n+1}}\left( 2\cdot\left\|\left\{A'\subseteq N ~\middle|~
       \sum_{j\in A'}2a_j=\alpha-2\right\}\right\|
   +\left\|\left\{A'\subseteq N ~\middle|~ \sum_{j\in
         A'}2a_j=\alpha-4\right\}\right\| \right).
 \]

 Allowing players with weight zero, there are different possibilities
 to split player $n+2$ into three players.  By
 Lemma~\ref{lem:bi-0-player}, splitting $n+2$ into one player with
 weight~$3$ and two others with weight~$0$ is not beneficial.
 Likewise, splitting $n+2$ into two players with weights~$1$ and~$2$
and one player with weight~$0$ is not beneficial, by
 Lemma~\ref{lem:bi-0-player} and since
 splitting into two players is not beneficial (by the remark above
 Theorem~\ref{thm:mbsb-pp-complete}).  Thus, the only possibility left
 is splitting $n+2$ into three players of weight~$1$ each.  This
 corresponds to the original game in the proof of
 Theorem~\ref{thm:bmb-pp-complete}, $\game_{i\div 3} =
 (2a_1,\dots,2a_n,1,1,1,1;\,\alpha)$.  Therefore,
\begin {eqnarray*}
\lefteqn{\bi(\game_{i\div 3},n+2)
 = \bi(\game_{i\div 3},n+3)
 = \bi(\game_{i\div 3},n+4)} \\
 & = & \frac{1}{2^{n+3}}\left(
    3\cdot\left\|\left\{A'\subseteq N ~\middle|~
 \sum_{j\in A'}2a_j=\alpha -2\right\}\right\|
  +\left\|\left\{A'\subseteq N ~\middle|~
 \sum_{j\in A'}2a_j=\alpha -4\right\}\right\|
    \right).
\end{eqnarray*}
 
Altogether, as in the proof of Theorem~\ref{thm:bmb-pp-complete}, the
sum of the three new players' probabilistic Banzhaf indices minus the
probabilistic Banzhaf index of the original player is greater than
zero if and only if
\[
 \left\|\left\{A'\subseteq N ~\middle|~ \sum_{j\in A'}a_j =
 \left(\nicefrac{\alpha}{2}\right)-1\right\}\right\| > 
 \left\|\left\{A'\subseteq N ~\middle|~ \sum_{j\in A'}a_j =
 \left(\nicefrac{\alpha}{2}\right)-2\right\}\right\|,
\]
which is true if and only if $A$ is in
\textsc{Compare}\hbox{-}\textsc{\#SubsetSum}\hbox{-}\textsc{\reflectbox{RR}}.

This result can be expanded to all $m\geq 3$ by splitting into
additional players with weight $0$. More precisely, if $m>3$, we
consider the same game $\game$ as above and split into three players
of weight~$1$ each and $m-3$ players of weight~$0$ each.  By
Lemma~\ref{lem:bi-0-player}, the sum of the $m$ new players' Banzhaf
power is equal to the combined Banzhaf power of the three players.
Thus, $\pp$-hardness holds by the same arguments as
above. %
\end{proofs}

On the other hand, a $\pp$ upper bound for the general beneficial
splitting problem cannot be shown in any straightforward way.  Here,
we can only show membership in~$\np^\pp$, and we conjecture that this
problem is even complete for this class.

\begin{theorem}
\label{thm:bsb-pp-hard-np-pp}
 $\bi$-\textsc{BeneficialSplit} is $\pp$-hard and belongs to~$\np^\pp$.
\end{theorem}

\begin{proofs}
  With $m$ being part of the input, there are exponentially many
  possibilities to distribute the split player's weight to her false
  identities.  Nondeterministically guessing such a distribution and
  then, for each distribution guessed, asking a $\pp$ oracle to check
  in polynomial time whether their combined Banzhaf power in the new
  game is greater than the original player's Banzhaf power in the
  original game, shows that $\bi$-\textsc{BeneficialSplit} is
  in~$\np^\pp$.

  Since $\bi$-$3$-\textsc{BeneficialSplit} is a special variant of the
  general problem $\bi$-\textsc{Bene}-\textsc{ficialSplit}, $\pp$-hardness is
  implied immediately by
  Theorem~\ref{thm:mbsb-pp-complete}.~\end{proofs}

\subsection{The Shapley--Shubik Power Index}
\label{sec:pp-completeness-shapley-shubik}

In order to prove $\pp$-hardness for the merging and splitting
problems with respect to the Shapley--Shubik index, we need to take a
further step back.

$\xthreeclong$ (\textsc{X3C}, for short) is another well-known
$\np$-complete decision problem: Given a set $B$ of size $3k$ and a
family $\mathscr{S}$ of subsets of $B$ that have size three each, does
there exist a subfamily $\mathscr{S}'$ of $\mathscr{S}$ such that $B$
is exactly covered by~$\mathscr{S}'$?

\begin{theorem}
\label{thm:bmsh-pp-complete}
 $\shi$-\textsc{BeneficialMerge} is $\pp$-complete, even if only two
players of equal weight merge.
\end{theorem}

\begin{proofs}
  The $\pp$ upper bound, which has already been observed for two
  players in~\cite{fal-hem:j:power-index-comparison}, can be shown
  analogously to the proof of Theorem~\ref{thm:bmb-pp-complete}.
 
  For proving the lower bound, observe that the size of a coalition a
  player is pivotal for is crucial for determining the player's
  Shapley--Shubik index.  Pursuing the techniques of Faliszewski and
  Hemaspaandra~\cite{fal-hem:j:power-index-comparison}, we examine the
  problem \textsc{COMPARE-\#X3C}, which is $\pp$-complete by
  Lemma~\ref{lem:compare-pp-complete}.  We will apply useful
  properties of \textsc{X3C} instances
  shown by Faliszewski and Hemaspaandra~\cite[Lemma
  2.7]{fal-hem:j:power-index-comparison}: Every \textsc{X3C} instance
  $(B',\mathscr{S}')$ can be transformed into an \textsc{X3C} instance
  $(B,\mathscr{S})$, where $\|B\| = 3k$ and $\|\mathscr{S}\| = n$,
  that satisfies $\nicefrac{k}{n}=\nicefrac{2}{3}$ without changing
  the number of solutions, i.e., $\#\textsc{X3C}(B,\mathscr{S}) =
  \#\textsc{X3C}(B',\mathscr{S}')$.  Now, by the properties of the
  standard reduction from \textsc{X3C} to \textsc{SubsetSum} (which in
  particular preserves the number of solutions, i.e.,
  \textsc{\#X3C} parsimoniously reduces to \textsc{\#SubsetSum},
  as well as the ``input size'' $n$ and the ``solution size''~$k$),
  we can assume that in a given \textsc{Compare-\#\subsetsum} instance
  each subsequence summing up to the given integer $q$ is of
  size~$\nicefrac{2n}{3}$.  Following the track of the reductions from
  \textsc{Compare-\#\subsetsum} via
  $\textsc{Compare}\hbox{-}\textsc{\#SubsetSum}\hbox{-}\textsc{R}$ to
  $\textsc{Compare}\hbox{-}\textsc{\#SubsetSum}\hbox{-}\textsc{RR}$ in
  Lemmas~\ref{lem:restr-compare-subsetsum-pp-complete}
  and~\ref{lem:even-more-restr-compare-subsetsum-pp-complete},
  a solution $A'\subseteq\{1,\dots,n\}$ to a given instance
  $A=(a_1,\dots,a_n)$ of the latter problem ($A'$ satisfying either
  $\sum_{i\in A'} a_i=(\nicefrac{\alpha}{2})-2$ or $\sum_{i\in
    A'} a_i=(\nicefrac{\alpha}{2})-1$, where $\alpha =\sum_{i=1}^n
  a_i$) can be assumed to satisfy $\|A'\| = k = \nicefrac{(n+2)}{3}$.
  Under this assumption, we show $\pp$-hardness of
  $\shi$-\textsc{BeneficialMerge} via a reduction from
  \textsc{Compare-\#\subsetsum-RR}.  Given such an instance, we
  construct the WVG $\game=(a_1,\dots,a_n,1,1;\,\nicefrac{\alpha}{2})$
  and consider coalition $S=\{n+1,n+2\}$.
  Define $X=\#\subsetsum(A,(\nicefrac{\alpha}{2})-1)$ and
  $Y=\#\subsetsum(A,(\nicefrac{\alpha}{2})-2)$.  Letting
  $N=\{1,\dots,n\}$, it holds that
 \begin{align*}
  &\shi(\game,n+1)=\shi(\game,n+2)\\&=
  \frac{1}{(n+2)!}\left(\left(\sum_{\substack{C\subseteq N \text{ such that}\\
  \sum\limits_{i\in C}a_i=(\nicefrac{\alpha}{2})-1}}
   \hspace{-3.5mm}\|C\|!(n+1-\|C\|)!\right)
   +\left(\sum_{\substack{C\subseteq N \text{ such that}\\
  \sum\limits_{i\in C}a_i=(\nicefrac{\alpha}{2})-2}}
   \hspace{-3.5mm}(\|C\|+1)!(n-\|C\|)!\right)\right)\\
  &=\frac{1}{(n+2)!}\left(
   X \cdot k!(n+1-k)! + Y \cdot (k+1)!(n-k)!\right).
 \end{align*}
 Merging the players in $S$, we obtain
 $\game_{\&S}=(2,a_1,\dots,a_n;\,\nicefrac{\alpha}{2})$.  The
 Shapley--Shubik index of the new player in $\game_{\&S}$ is
 \begin{align*}
 \shi(\game_{\&S},1)&=
  \frac{1}{(n+1)!}\sum_{\substack{C\subseteq N  \text{ such that}\\
  \sum\limits_{i\in C}a_i\,\in\,\{(\nicefrac{\alpha}{2})-1,(\nicefrac{\alpha}{2})-2\}}}
   \hspace{-3mm}\|C\|!(n-\|C\|)!\\
  &=\frac{1}{(n+1)!}\left(
   X + Y \right) \cdot (k+1)!(n-k)!.
 \end{align*}
 All in all,
 \begin{align}
  &\shi(\game_{\&S},1)-\left(\shi(\game,n+1) + \shi(\game,n+2)\right)
 \nonumber\\
  &=\frac{\left(X + Y \right) \cdot (k+1)!(n-k)!}{(n+1)!}
  -\frac{2\left(X \cdot k!(n+1-k)! + Y \cdot (k+1)!(n-k)!\right)}{(n+2)!}\nonumber\\
 &=\frac{k!(n-k)!}{(n+2)!}(n-2k)(-X+Y) \label{eq:sh}.
 \end{align}
 Since we assumed that $k=\nicefrac{(n+2)}{3}$ and we can also assume that
 $n>4$ (because we added four integers in the construction in the proof of
 Lemma~\ref{lem:even-more-restr-compare-subsetsum-pp-complete}),
 it holds that
 \[
  n-2k=\frac{n-4}{3}>0.
 \]
 Thus the term (\ref{eq:sh}) is greater than zero if and only if
 $Y$ is greater than~$X$, which is true if and only if
 $A$ is in \textsc{Compare-\#SubsetSum-RR}.~\end{proofs}

\begin{theorem}
\label{thm:mbssh-pp-complete}
 $\shi$-$m$-\textsc{BeneficialSplit} is $\pp$-complete for each $m \geq 2$.
\end{theorem}

\begin{proofs}
  $\pp$ membership can be shown analogously to the $\pp$ upper bound
  in the proof of Theorem~\ref{thm:mbsb-pp-complete}.
  $\pp$-hardness can also be shown
  analogously to the proof of Theorem~\ref{thm:mbsb-pp-complete},
  appropriately modified to use the arguments from the proof of
  Theorem~\ref{thm:bmsh-pp-complete} instead of those from the proof
  of Theorem~\ref{thm:bmb-pp-complete}.~\end{proofs}

\begin{theorem}
\label{thm:bssh-pp-hard-np-pp}
 $\shi$-\textsc{BeneficialSplit} is $\pp$-hard and belongs to $\np^\pp$.
\end{theorem}

\begin{proofs}
  The upper bound of $\np^\pp$ holds due to analogous arguments as in
  the proof of Theorem~\ref{thm:bsb-pp-hard-np-pp}.  Also, analogously
  to the proof of Theorem~\ref{thm:bsb-pp-hard-np-pp}, since
  $\shi$-$2$-\textsc{Beneficial}-{Split} is a special variant of the
  general $\shi$-\textsc{Beneficial-Split} problem, $\pp$-hardness is
  implied immediately by
  Theorem~\ref{thm:mbssh-pp-complete}.~\end{proofs}

\section{Conclusions and Open Questions}
\label{sec:conclusions}

Solving previous conjectures in the affirmative, we have pinpointed
the precise complexity of the beneficial merging problem in weighted
voting games for the Shapley--Shubik and the probabilistic Banzhaf
index by showing that it is $\pp$-complete.  We have obtained the same
result for beneficial splitting (a.k.a.\ false-name manipulation)
whenever the number of false identities a player splits into is fixed.
For an unbounded number of false identities, we raised the known lower
bound from $\np$-hardness to $\pp$-hardness and showed that it is
contained in $\np^{\pp}$.  For this problem, it remains open whether
it can be shown to be complete for $\np^{\pp}$, a huge complexity
class that by Toda's theorem~\cite{tod:j:pp-ph} contains the entire
polynomial hierarchy.  $\np^{\pp}$ is an interesting class, but
somewhat sparse in natural complete problems.  The only (natural)
$\np^{\pp}$-completeness results we are aware of are due to
Littman et
al.~\cite{lit-gol-mun:j:complexity-probabilistic-planning}, who
analyze a variant of the
satisfiability problem and questions related to probabilistic planning,
and due to Mundhenk
et
al.~\cite{mun-gol-lus-all:j:finite-horizon-markov-decision-process-complexity},
who study problems related to finite-horizon Markov decision processes.

Another interesting open question is whether our results can be
transferred also to the beneficial merging and splitting problems
for the normalized Banzhaf index.
Finally, it would be interesting to know to which classes of simple
games, other than weighted voting games, our results can be extended.

\noindent
\paragraph{Acknowledgments:}
We thank Haris Aziz for interesting discussions.
This work has been supported in part by DFG grant RO-1202/14-1.

%
%
%
% \bibliography{games}
\bibliographystyle{alpha}
\newcommand{\etalchar}[1]{$^{#1}$}

\end{document}